\documentclass[12pt,a4paper,final]{iopart}

\usepackage{graphicx}
\usepackage{dcolumn}
\usepackage{bm}

\expandafter\let\csname equation*\endcsname\relax
 \expandafter\let\csname endequation*\endcsname\relax
\usepackage{amsmath}
\usepackage{upgreek}
\usepackage{physics}
\usepackage{isomath}
\usepackage{placeins}
\newcommand{\pd}[2]{\frac{\partial #1}{\partial #2}}

\begin{document}

\title{Driving Quantum Correlated Atom-Pairs from a Bose-Einstein
  Condensate}

\author{Liang-Ying Chih}
\address{JILA and
  Department of Physics, University of Colorado, Boulder, Colorado
  80309-0440, USA.}
\author{Murray Holland} 
\address{JILA and
  Department of Physics, University of Colorado, Boulder, Colorado
  80309-0440, USA.}
\date{\today}

\begin{abstract}
  The ability to cool quantum gases into the quantum degenerate realm
  has opened up possibilities for an extreme level of quantum-state
  control. In this paper, we investigate one such control protocol
  that demonstrates the resonant amplification of quasimomentum pairs
  from a Bose-Einstein condensate by the periodic modulation of
  the two-body $s$-wave scattering length. This shows a capability to
  selectively amplify quantum fluctuations with a predetermined
  momentum, where the momentum value can be spectroscopically
  tuned. A classical external field that excites pairs of particles
  with the same energy but opposite momenta is reminiscent of the coherently-driven nonlinearity in a
  parametric amplifier crystal in nonlinear optics. For this reason,
  it may be anticipated that the evolution will generate a `squeezed'
  matter-wave state in the quasiparticle mode on resonance with the
  modulation frequency. Our model and analysis is motivated by a
  recent experiment by Clark {\em et al.} that observed a
  time-of-flight pattern similar to an exploding
  firework~\cite{Firework}. Since the drive is a highly coherent
  process, we interpret the observed firework patterns as arising from
  a monotonic growth in the two-body correlation amplitude, so that
  the jets should contain correlated atom pairs with nearly equal and
  opposite momenta. We propose a potential future experiment based on
  applying Ramsey interferometry to experimentally probe these pair
  correlations.
\end{abstract}

\maketitle

\section{\label{sec:intro}Introduction}
The ability to tune the two-body scattering length in a Bose-Einstein
Condensate (BEC) by varying the magnitude of a magnetic field in the
vicinity of a Feshbach resonance has been employed in a number of
seminal experiments that aim to investigate controlled non-equilibrium
quantum dynamics. One such example is the so-called `Bosenova'
experiment by Donley {\em et al.}~\cite{Donley2001}, in which a BEC
was subject to a sudden change of the scattering length from a small
positive value to a large negative value. This resulted in a change of
the sign of the mean-field interactions from repulsive, where the gas
is mechanically stable, to attractive, where the compressibility may be
negative and the gas is then unstable~\cite{Ruprecht}. What was observed
experimentally after this abrupt change in the scattering length was a
collapse and subsequent explosion of the quantum gas in a manner that
resembled an astrophysical supernova. Theoretical models were
subsequently developed and illustrated that the emergence of a
pairing field in the underlying quantum many-body system can explain
the observed burst of non-condensate atoms~\cite{Milstein_2003}.

More recently, Bose `firework' experiments~\cite{Firework} have
observed pairs of high momentum atoms emitted as jets from a
condensate driven by a periodic modulation of the two-body $s$-wave
scattering length. These experiments demonstrated a protocol for
resonantly amplifying quantum fluctuations with well-controlled
momenta when starting from a stationary BEC. The fact that the jets
were observed to be correlated in emission direction motivates us to
consider whether the many-body pairing field played an important
role in the dynamics, in a similar manner to the Bosenova system
previously studied. This aspect is related to other calculations that explore
the second order coherence of the gas~\cite{Theory3}.

In the case of a dilute quantum gas, the Gross-Pitaevskii equation
(GPE) provides an accurate description of the equilibrium and
time-dependent behavior of the BEC. In this framework, the interacting
condensate is completely described by a mean-field superfluid order
parameter. The GPE framework has been extensively applied to model the
behavior of BECs at zero temperature, and also to their coherent
manipulation through externally applied potentials. However, when
there is a significant portion of non-condensate atoms, the GPE will
fail to provide an accurate description of the system. A small amount
of non-condensate atoms is always present even at zero temperature in
a dilute quantum gas arising from the beyond mean-field fluctuations
that are due to the finite interaction strength. It is of course
possible to generate substantial fractions of non-condensate atoms by driving a pure condensate in
a variety of ways, and this is typically unavoidable when the currents that generate the magnetic confinement fields contain stochastic noise. Furthermore, non-condensate atoms are
always present in systems with finite temperature since they embody the thermal excitations. In order to capture the essential dynamics
associated with the non-condensed component, a theory that goes beyond
the mean-field approximation is necessary.

A systematic extension to the simplest mean-field approach given by
the GPE is the Hartree-Fock Bogoliubov (HFB) formalism that takes into
account the interactions between three components. We will refer
to these as the condensate, the non-condensate, and the pairing field of the
fluctuations. In this formalism, the elementary excitations are
described as Bogoliubov quasiparticles \cite{Superfluidity}, and the
ground state condensate is a vacuum of such quasiparticles. The
quasiparticle creation operator is a linear combination of the atom
creation operator (particles) and the atom annihilation operator
(holes). The vacuum state due to interactions possesses a portion of
non-condensate atoms referred to as quantum depletion.  The recent
`firework' experiments that tune the scattering length by applying an
appropriate external magnetic field potentially allow all of these
components---the mean-field, non-condensate, and pairing field---to be
controlled, manipulated, and engineered. In this paper we derive the
solutions of the HFB theory as applied to well-controlled
experimental geometries in order to determine the efficacy of this
framework for providing a theoretical basis for the recent
observations.

One approach for describing collective excitations of a condensate
involves solving the Bogoliubov de Gennes equations~\cite{Theory2},
which is most accurate when the excitations are weak. When the
excitations are not weak, and the non-condensate fraction can be
significant, a more complete approach must be used such as the time-dependent
self-consistent HFB equations~\cite{Theory1}, and that is the method
 we will focus on in this paper. Note that one alternative approach
that can incorporate the excitations is through the addition of a
noise source to introduce fluctuations directly into the
Gross-Pitaevskii equation~\cite{DensityWaves}. However, this assumes by construction that the many-body state can be
accurately described by a unique macroscopic wavefunction, and
therefore a more complete theory is needed to describe two-body correlations.

We emphasize the importance of the pairing field in our
analysis. Pairing gives rise to an anomalous density that allows us to
investigate the coherence of the system and explore methods to probe
phase-sensitive quantities. However, to incorporate the pairing field
in our simulations requires a number of important
considerations. Since our modeling assumes contact interactions,
numerical studies have to account for the potentially divergent nature
of the pairing field at both short and long length scales by
appropriate renormalization of the scattering potential. We
demonstrate how to renormalize the scattering potential when momentum
is represented on a discrete grid. Furthermore, when solving for the
initial condition of the system, instead of using an approximation
that ignores the pairing field~\cite{Popov} in order to remedy issues
associated with the gapless energy spectrum, we take an alternative approach in which the condensate,
depletion and pairing field are accounted for and we solve for the HFB theory
self-consistently. 

Note that there is one other important consideration; our model does
not include the collisions terms in the kinetic theory that result in equilibration of the gas to its thermal state \cite{KineticTheory}.  Neglecting collisions is a good approximation
for a dilute gas at low temperature, but implicitly requires us to
limit our discussion to the regime in which the time-scale between
consecutive two-body collision events greatly exceeds the time-scale
of the quantum dynamics that we investigate. Finally, all this has to be
implemented in multiple dimensions in order to provide a useful
comparison with experimental observations.

The paper is outlined as follows. We present the model in
Section~\ref{sec:fieldtheory} and provide details of the
renormalization process in Section~\ref{sec:renormalization}. At
first we limit our discussion to the most straightforward case of
quasi-1D systems. In Section~\ref{sec:HFB}, we outline the numerical
procedures necessary to obtain a self-consistent ground state solution
to the HFB theory for a weakly-interacting trapped quantum gas, and
quantify the quantum depletion as well as the pairing field
amplitude. In Sections~\ref{sec:TDHFB} and~\ref{sec:squeezing}, we use
the time-dependent HFB theory to show that the modification of the
interaction strength through modulation of the scattering length
parametrically amplifies a certain quasiparticle mode and generates a
matter-wave solution that is analogous to a squeezed state of light.
In Section~\ref{sec:interference}, we use these results to explore the
possibility of future experiments that utilize interferometry to probe
the pair correlation amplitude. We consider two methods that create a
phase difference between the driving field and the pairing field, and
consequently lead to the possibility for constructive and destructive
interference in the matter-wave density. Finally, in
Section~\ref{sec:2D}, we extend the results to quasi-2D so that they
can be compared with the experimental observations of angular
correlations in the firework pattern, where the atoms were confined in a pancake-shaped
confining potential well and were ballistically expanded.

\section{General many-body field theory}
\label{sec:fieldtheory}
We begin from the many-body Hamiltonian that describes a weakly
interacting Bose gas with pairwise contact interactions:
\begin{eqnarray}
  \mathcal{H} & = &\int d^3x\,\hat{\psi}^\dagger
                    (\vectorsym{x})\left(-\frac{\hbar^2}{2m}
                    \nabla^2+V_\text{ext}
                    (\vectorsym{x})\right)
                    \hat{\psi}(\vectorsym{x})\nonumber\\
              &&{}\quad
                 +\frac{V}{2}\int d^3x\,\hat{\psi}^\dagger
                 (\vectorsym{x})\hat{\psi}^\dagger(\vectorsym{x})
                 \hat{\psi}(\vectorsym{x})
                 \hat{\psi}(\vectorsym{x})\,,
    \label{eq:hamiltonian}
\end{eqnarray}
where $m$ is the mass of the atom and $V_{\rm ext}$ is the external
trapping potential. The field operators, $\hat{\psi}(\vectorsym{x})$ and
$\hat{\psi}^\dagger(\vectorsym{x})$, are bosonic operators that
annihilate and create particles and obey commutation relations
$[\hat{\psi}(\vectorsym{x}),\, \hat{\psi}^\dagger(\vectorsym{x}')]=
\delta(\vectorsym{x}-\vectorsym{x}')$. The strength of the interaction
potential,~$V$, is related to the $s$-wave scattering length,~$a$, by
$V=T\Gamma$, where $T=4\pi\hbar^2a/m$ is the
three-dimensional $T$-matrix (here it is actually a simple scalar and not a matrix
since we consider the regime in which there is no dependence of the scattering phase shift on energy)
and $\Gamma$ is the dimensionless renormalization factor that will be fully discussed in
Section~\ref{sec:renormalization}. 

Since we intend to explore excitations from a BEC, we assume that the
field operator is well described by a mean field amplitude describing
the atom condensate, $\phi_a(\vectorsym{x})$, and a fluctuating
component, i.e.,
\begin{equation}
  \hat{\psi}(\vectorsym{x}) = \langle\hat{\psi}(\vectorsym{x})\rangle
  + \delta\hat{\psi}(\vectorsym{x}) = \phi_a(\vectorsym{x})
  + \delta\hat{\psi}(\vectorsym{x})\,.
\end{equation}
where $ \langle\delta\hat{\psi}(\vectorsym{x})\rangle=0$. The
second-order terms---normal and anomalous densities---are defined
respectively as,
\begin{eqnarray}
  G_N(\vectorsym{x}, \vectorsym{x}')
  &=& \langle \delta\hat{\psi}^\dagger(\vectorsym{x}') \delta\hat{\psi}
      (\vectorsym{x})\rangle\,,\nonumber\\
  G_A(\vectorsym{x},
  \vectorsym{x}')
  &=& \langle \delta\hat{\psi}(\vectorsym{x}')
      \delta\hat{\psi}(\vectorsym{x}) \rangle\,.
\end{eqnarray}
Both of these play an important role in the dynamics of the
non-condensate component of the system we are interested in. In particular, the
diagonal elements of the normal density, $G_N(\vectorsym{x}, \vectorsym{x})$, represent the physical
non-condensate atom densities at position $\vectorsym{x}$ and are
therefore positive semi-definite. The off-diagonal elements represent the matter-wave
correlations of the non-condensate atoms that are characterized by
quantities such as the de Broglie wavelength and effective
temperature. The anomalous density, $G_A(\vectorsym{x}, \vectorsym{x})$, is the pairing field that
characterizes the two-particle correlations in the system.

If we assume that the the field fluctuations are Gaussian, one can
drop the third-order cumulants, and expand the fourth-order quantities
in terms of the second-order cumulants when deriving the evolution
equations. In practice, this involves repeated application of Wick's
theorem~\cite{Peskin_Schroeder}. The resulting equations of motion are
closed and can be written in detail for the condensate as;
\begin{eqnarray}
  i\hbar\pd{\phi_a(\vectorsym{x})}{t}
  &=&
      \left(-\frac{\hbar^2}{2m}\nabla^2+V_\text{ext}
      (\vectorsym{x})\right)\phi_a(\vectorsym{x})
      \nonumber\\
  &&\quad{}
     +V\left[|\phi_a(\vectorsym{x})|^2+ 2G_N(\vectorsym{x},
     \vectorsym{x})\right]\phi_a(\vectorsym{x})
     \nonumber\\
  &&\quad{}+V\,G_A(\vectorsym{x},\vectorsym{x})\phi_a^*
     (\vectorsym{x})\,,
     \label{eq:eom_phiA}
\end{eqnarray}
for the normal density as;
\begin{eqnarray}
  i\hbar \pd{}{t}G_N(\vectorsym{x},\vectorsym{x'})
  &=&
      \mathcal{H}'(\vectorsym{x})G_N(\vectorsym{x},\vectorsym{x'})
      - \mathcal{H}'(\vectorsym{x'})
      G_N(\vectorsym{x},\vectorsym{x'})\nonumber\\
  &&\,+\Delta(\vectorsym{x})
     G_A^*(\vectorsym{x},\vectorsym{x'})-\Delta^*
     (\vectorsym{x'})G_A(\vectorsym{x},\vectorsym{x'})
     \label{eq:eom_GN}
\end{eqnarray}
and for the anomalous density as;
\begin{eqnarray}
  i\hbar \pd{}{t}G_A(\vectorsym{x},\vectorsym{x'})
  &&= \mathcal{H}'(\vectorsym{x})G_A(\vectorsym{x},\vectorsym{x'})
     + \mathcal{H}'(\vectorsym{x'})G_A(\vectorsym{x},\vectorsym{x'})
     \nonumber\\
  &&\quad{}+\Delta(\vectorsym{x})\left[
     G_N^*(\vectorsym{x},\vectorsym{x'})
     +\delta(\vectorsym{x}-\vectorsym{x'})\right]
     \nonumber\\
  &&\quad{}+\Delta^*(\vectorsym{x'})G_N(\vectorsym{x},\vectorsym{x'})\,.
\label{eq:eom_GA}
\end{eqnarray}
Here we have simplified the notation by introducing two energy
functionals,
\begin{eqnarray}
  \mathcal{H}'(\vectorsym{x})
  &=& -\frac{\hbar^2}{2m}\nabla^2+V_\text{ext}(\vectorsym{x})
      +2V\left[|\phi_a(\vectorsym{x})|^2
      + G_N(\vectorsym{x}, \vectorsym{x})\right]\,,\nonumber\\
  \Delta(\vectorsym{x})
  &=& V\left[\phi_a(\vectorsym{x})^2+G_A(\vectorsym{x},
      \vectorsym{x})\right]\,,
      \label{eq:numberAndGap}
\end{eqnarray}
for the single-particle self-energy and the gap, respectively. Due to
the fact that we neglect explicit three-particle and higher
correlations, the validity of this approach is restricted to the
dilute gas regime. Note that equation~(\ref{eq:eom_phiA}) can simplified to
the Gross-Pitaevskii equation if terms involving
$G_N(\vectorsym{x},\vectorsym{x})$ and
$G_A(\vectorsym{x},\vectorsym{x})$ are dropped. In this case, the
time-independent energy eigenvalue represents the chemical potential,
$\mu$, so that $i\hbar\dot\phi_a=\mu\phi_a$. The delta function in
equation~(\ref{eq:eom_GA}) arises from the bosonic commutation relation of
the field operators and can therefore be interpreted as a quantum effect. A number of
quantities are conserved in this evolution; in particular, the total
atom number
\begin{equation}
  \mathcal{N} = \int d\vectorsym{x}\left(|\phi_a(\vectorsym{x})|^2
  +G_N(\vectorsym{x}, \vectorsym{x})\right)
\end{equation}
is invariant under time evolution governed by
equations~(\ref{eq:eom_phiA})--(\ref{eq:numberAndGap}).

In order to see how the anomalous density is related to the vacuum pair
wavefunction for the interatomic separation of
two atoms, we may neglect the mean-field density and the normal density
in equation~(\ref{eq:eom_GA}), and then the eigenvalue equation is simplified to
\begin{equation}
  -\frac{\hbar^2}{m}\nabla^2 G_A(\vectorsym{r})+V\delta(\vectorsym{r})
  G_A(\vectorsym{r}) = 2\mu G_A(\vectorsym{r})\,,
    \label{eq:scattering}
\end{equation}
where
$\vectorsym{r}=\vectorsym{x}-\vectorsym{x}'$. equation~(\ref{eq:scattering})
can be identified as a one-dimensional Schr\"odinger equation of a
fictitious particle of reduced mass $m/2$ scattering off a potential
$V\delta(\vectorsym{r})$. Then $G_A(\vectorsym{r})$ is interpreted
as the resulting eigenstate wavefunction corresponding to the familiar two-particle scattering solution of the equation written in terms of the relative
coordinate.

\section{Renormalization of the Scattering Potential}
\label{sec:renormalization}
The Dirac delta function in equation~(\ref{eq:scattering}) implies that we
are implicitly building a scattering model from a contact
interaction. This is convenient as it simplifies the resulting field
theory, but care must be taken to account for divergences that can
arise at small and large scales. In general, this is remedied by renormalization
of the potential strength. In order to carry out this renormalization procedure, we begin from the
formal scattering theory~\cite{Taylor_scattering}, where we define the
bare scattering potential operator, $\hat{V}$, which has units of
energy, and thereby expand the $T$-matrix in an order-by-order series;
\begin{eqnarray}
  \hat{T} &=& \hat{V} + \hat{V}G_0\hat{V} + \hat{V}G_0\hat{V}G_0\hat{V}
             + \ldots\nonumber\\
          &=& \hat{V} + \hat{V}G_0\hat{T}\,.
              \label{eq:scatteringseries}
\end{eqnarray}
Here $G_0$ is the bare single particle propagator,
\begin{equation}
  G_0 = \frac{1}{E-\hat{H}_0+i\epsilon}\,,
\end{equation}
the scattering energy is $E$, the dispersion relation is
$\hat{H}_0=\hat{p}^2/(2m)$ with $\hat{p}$ the momentum, and we need to
implicitly consider the limit $\epsilon\rightarrow0$. The $T$-matrix elements are
$T=\left\langle\vectorsym{k}'\right\lvert\hat{T}\lvert\vectorsym{k}\rangle$, where
$\ket{\vectorsym{k}}$ is the wavenumber basis state. For the low
energy scattering limit, the $T$-matrix becomes independent of $E$,
and does not depend on $\vectorsym{k}$ or $\vectorsym{k'}$. In this
case the $T$-matrix is well characterized by a constant scalar
associated with the $s$-wave scattering length, as mentioned earlier, i.e.,
$T=4\pi\hbar^2a/m$.

Further considerations have to be made when one or more dimensions are effectively
frozen out due to imposing a strong confining potential in these dimensions. Without loss
of generality, let us consider the strong confining potential to be a
harmonic potential with oscillator length given by $l_{\perp}$. If one
dimension is frozen out, an effective quasi-2D geometry is realized,
and if two dimensions are frozen out, an effective quasi-1D system is
generated. If we denote the number of free dimensions by
$n\in\{1,2,3\}$, the appropriate $T$-matrix expression, $T_n$, for the
reduced dimensional case can be related recursively by $T_3=T$ and
$T_{n-1}=T_n/(\sqrt{2\pi}l_{\perp})$~\cite{Olshanii_2003}\cite{2D_Tmatrix}.

The process of renormalization connects the $T$-matrix, $T_n$, to the
strength of the potential, $V_n$, by expanding
equation~(\ref{eq:scatteringseries}) in the momentum basis, and this
connection depends on the dimensionality of the system,
\begin{equation}
  T_n=V_n+V_n\int_{K_-}^{K_+} \frac{d^nk}{(2\pi)^n}
  \frac{T_n}{E-\frac{\hbar^2k^2}{2m}}\,,
  \label{eq:renorm}
\end{equation}
where the critical element here is the introduction of $K_-$ and $K_+$
as infrared and ultraviolet momentum cutoffs, respectively. The
cutoffs have to be chosen from an appropriate asymptotic limit in
order to accurately capture the dynamics of interest. The renormalization procedure can be represented by the introduction of a parameter,
$\Gamma_n$, defined by solving equation~(\ref{eq:renorm}) for $V_n$. This
gives the solution,
\begin{equation}
V_n = \frac{T_n}{1-\alpha_n T_n} \equiv
T_n\Gamma_n\,,
\end{equation}
where
\begin{equation}
  \alpha_n = -\int_{K_-}^{K_+} \frac{d^nk}{(2\pi)^n}
  \frac{1}{E-\frac{\hbar^2k^2}{2m}}\,.
\end{equation}
In order to illustrate the behavior of $\alpha_n$, we consider the
solution to scattering equation, equation~(\ref{eq:scattering}), where the
stationary energy eigenvalue is $E=2\mu$ and the mass is replaced by
the reduced mass of two particles, $m\to m/2$.  Solving this system of
equations has a character that depends on the dimensionality. In three
dimensions, we may set $K_-=0$, and perform the integral to give
$\alpha_3=mK_+/(2\pi^2\hbar^2)$ for a particle scattering at low
energy, $E\rightarrow0$~\cite{renormalization}. In 2D, the integral
scales logarithmically and has both ultraviolet and infrared
divergences. In 1D, there is an infrared divergence so $K_-$ must be
non-zero but we may set $K_+$ to infinity.

We do not provide all the details here, since, in practice, these are formal
considerations that do not actually affect our numerical simulations. Indeed there are actually no
 divergences introduced that require the introduction of momentum cutoffs to rectify when the momenta are restricted to values on
discrete and finite grids. This is always the case in a numerical computer model that aims to describe a realistic experiment. In such a discrete
representation of possible momenta, it is preferable to simply calculate a finite sum over a specific partition instead of evaluating the continuous integral analytically. This
implies a numerical evaluation of
$\sum_{i=0}^{N-1}\left(\hbar^2k_i^2/m-2\mu\right)^{-1}$, giving
$\alpha_n$, and therefore determining $V_n$ for a given $T$-matrix,
which replaces $V$ in the HFB equations, i.e.,
equations~(\ref{eq:eom_phiA})--(\ref{eq:eom_GA}). Here the subscripts $i$
label individual discrete momenta, and thus $\{k_i\}$ represents the
momentum grid, with $N$ is the total number of grid points.

We carry out this renormalization procedure for all the results that
we present in this paper. For each calculation, we verify that
the numerical results are independent of the details of the momentum grid on
which the field theory is represented.

\section{Self-consistent ground state solution}
\label{sec:HFB}
In order to find a self-consistent solution to prepare an initial
condition for the subsequent time evolution, the first step will be to
consider the non-condensate component to be absent, and to find a
ground state representation of the condensate by solving the GPE. We then
use this condensate field as input into the time-independent equations
for the normal and anomalous densities, and diagonalize the resulting
HFB self-energy matrix to find the quasiparticle basis. As we will
see, this process exhibits a defect in the zero-energy subspace (i.e., the
eigenvectors do not span the space). The intepretation is that the
eigensolution is not stationary and cannot be used as an accurate
description of the initial condition for subsequent time evolution. We
therefore reintroduce the non-condensate terms that we have just found
into the equations for the condensate, normal density, and anomalous
density and solve again the system of equations, giving rise to an iterative method that generates
an accurate self-consistent initial condition.

Our approach will be to begin by first fully describing the necessary procedure using the
simple case of quasi-1D where the problem is most easily tractable. However,
higher dimensions can be treated in a similar method to the manner we
present (we will consider quasi-2D later in Section~\ref{sec:2D}). The
reduction to one-dimensional behavior requires the transverse confinement condition
\begin{equation}
  \frac{a}{n_\text{1D}l_{\perp}^2} \ll 1
\end{equation}
to be satisfied, where $n_{\rm 1D}$ is the one-dimensional
density~\cite{Olshanii_1998}\cite{Lieb_Liniger}, and as defined
previously, $l_{\perp}$ is the harmonic oscillator length in the two
strongly confining directions, here assumed to be equal.

The first part of our numerical algorithm is to solve for the ground
state of the GPE,
\begin{equation}
  i\hbar\pd{\phi_a(x)}{t}
  =
      \left(-\frac{\hbar^2}{2m}\nabla^2+V_\text{ext}
      (x)\right)\phi_a(x)
    +T_1|\phi_a(x)|^2\phi_a(x)\,.
    \label{eq:1dgpe}
\end{equation}
We use imaginary-time propagation to derive the lowest energy solution
and its energy eigenvalue $\mu$ representing the associated chemical
potential. Then, the mean field solution, $\phi_a(x)$, can be used as a
parameter to construct the self-energy matrix;
\begin{equation}
    \Sigma = 
    \begin{pmatrix}
    \Sigma_N & \Sigma_A\\
    -\Sigma_A^* & -\Sigma_N^*
  \end{pmatrix}\,,
  \label{eq:simplesigma}
\end{equation}
where
\begin{eqnarray}
  \Sigma_N &=& -\frac{\hbar^2}{2m}\nabla^2
               - \mu+2T_1|\phi_a(x)|^2\,,\nonumber\\
  \Sigma_A &=& T_1\phi_a(x)^2\,.
  \label{eq:sigmas}
\end{eqnarray}
The self-energy matrix has
dimensionality $2N\times 2N$ where $N$ is the size of the single
particle basis, as defined previously. This energy operator is most simply expressed in the position basis, where $x\in(0, L]$, since in that representation the potential terms including the mean-field appear as diagonal blocks. The eigenstates of $\Sigma$ are the Bogoliubov
quasiparticles. 
Since the matrix satisfies $\sigma_z\Sigma=\Sigma^\dagger\sigma_z$, where $\sigma_z =\text{diag}(I_{N\times N},\,-I_{N\times N})$, 
the eigenenergies come in pairs of positive and negative values,~$\pm \epsilon_k$, and the corresponding eigenstates are
$w_k=(u_k(x), v_k(x))^T$ and $w_{-k}=(v_k^*(x), u_k^*(x))^T$. The
eigenstates are normalized by satisfying the constraint
\begin{equation}
    \int _0^L dx\left(|u_k(x)|^2-|v_k(x)|^2\right)\longrightarrow
    \sum_{j=1}^N\frac{L}{N}\left(|u_k(jL/N)|^2-|v_k(jL/N)|^2\right)=1\,,
    \label{eq:normalization}
\end{equation}
Although this construction may appear standard and straightforward, there is a well-known and implicit subtlety
when examining the solutions to this eigensystem. When investigating
the zero-energy eigensolutions, one finds a pair of eigenstates that
are colinear (equal up to a multiplicative scalar) that have the form
$\mathcal{P}=(\phi_a(x)/\sqrt{2\mathcal{N}}, -\phi_a^*(x)/\sqrt{2\mathcal{N}})^T$. This
solution creates two significant issues. First, the colinear eigenstates cannot be normalized by
equation~(\ref{eq:normalization}). Second, they do not span the two-dimensional subspace of
the Hilbert space corresponding to zero-energy.

The origin of this mathematical fact has an intuitive
explanation. It arises from the approximations that
lead to this self-energy matrix, that is, by fixing the condensate
solution as an unchanging parameter, one builds an unphysical model that
implicitly allows the unconstrained growth of a zero energy mode as
a function of increasing time. Consequently there is no stationary
solution. This has to be remedied, for example, through a
self-consistent approach in which the condensate is treated as a variational parameter, in order to allow us to extend the
formalism so that it may be applied to our system of interest.

We begin by determining the remaining eigenvector to fully span the
zero-energy subspace by employing the Gram-Schmidt orthogonalization
method to numerically calculate the remaining basis vector. In this way we determine an eigenvector solution
$\mathcal{Q}=(q(x), -q^*(x))^T$ such that
$\frac{L}{N}\mathcal{Q}^\dagger\sigma_zw_k=0$ for all $k\neq 0$, and normalize it
to $\frac{L}{N}\mathcal{Q}^\dagger\sigma_z\mathcal{P}=i$
\cite{WALSER2004107}. 
The addition of this vector to the eigenvectors
of the self-energy completes the basis of the vector space. The reason that this is
important is that it allows the field operator to be expanded as
\begin{equation}
  \delta\hat{\psi}(x) = \sum_{k=1}^{N-1}(u_k(x)\hat{b}_k+v_k^*(x)\hat{b}_k^\dagger)
  -i\frac{\phi_a(x)}{\sqrt{2\mathcal{N}}}\hat{\theta}+iq(x)\hat{L}
\end{equation}
where $\hat{b}_k$ and $\hat{b}_k^\dagger$ are bosonic annihilation and
creation operators for the quasiparticles. We have introduced
$\hat{\theta}$ and $\hat{L}$ as a pair of canonically conjugate
operators that fully describe the zero-energy mode and obey the cannonical commutation relation $[\hat{\theta}$,\
$\hat{L}]=i$.

It is convenient to identify two special combinations of $\mathcal{P}$
and $\mathcal{Q}$ in order to give a concise expression for the
completeness relation. We define
$w_\pm=(\mp\mathcal{P}+i\mathcal{Q})/\sqrt{2}$, along with the matrix
\begin{equation}
    W=\begin{pmatrix}
    w_+, w_1, ...,w_{N-1}, w_-, w_{-1}, ..., w_{-(N-1)}
  \end{pmatrix}
  \label{eq:ws}
\end{equation}
so that the following completeness relation is satisfied;
\begin{align}
    \frac{L}{N}W^\dagger \sigma_z W = \sigma_z\,.
\end{align} 
This allows the particle
annihilation operator to be written as
\begin{equation}
  \delta\hat{\psi}(x) = \sum_{k\in{\cal
      S}}(u_k(x)\hat{b}_k+v_k^*(x)\hat{b}_k^\dagger)\,,
\end{equation}
where the sum is over the elements of the index set
${\cal S}=\{+,1,2,\ldots,N-1\}$, and $\hat{b}_+$ is the annihilation operator for the zero-energy mode given by $\hat{b}_+=(i\hat{\theta}+\hat{L})/\sqrt{2}$.

At this point, we have determined the quasiparticle basis, and can
populate that basis with a given set of probabilities in order to
generate particle distributions. In particular, we would like to
derive the normal $G_N(x,x')$ and anomalous $G_A(x,x')$ densities that
are essential elements of the HFB theory. To begin with we construct
the Hermitian density matrix:
\begin{eqnarray}
    G&=&\begin{pmatrix}
    G_N(x,x') & G_A(x,x')\\
    G_A^*(x,x') & \delta(x-x')+G_N^*(x,x')\end{pmatrix}\nonumber\\
    &=&W\Pi W^\dagger,
    \label{eq:factorizeG}
\end{eqnarray}
where the population matrix $\Pi$ has the form
\begin{equation}
   \Pi=\begin{pmatrix}
    p & q\\
    q^* & I+p
  \end{pmatrix}.
  \label{eq:pq_matrix}
\end{equation}
The diagonal elements of $p$ are the populations of each quasiparticle
$\langle \hat{b}_k^\dagger \hat{b}_k\rangle$, and the off-diagonal elements
represent the correlations between different quasiparticles. In the
ground state, $p=0$ and $q=0$. The identity, $I$, on the lower-right
block is interpreted as a bosonic analog to the Dirac sea \cite{DiracSea}, in which
the negative energy states are occupied by boson holes. When there is
an excitation, a pair of one particle and one hole is created, and
therefore $p$ appears in both the upper-left and the lower-right
block, as shown in equation~(\ref{eq:pq_matrix}).

\begin{figure}
  \centering \includegraphics[width=0.4\textwidth]{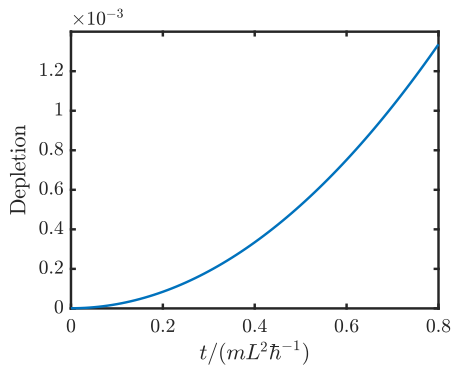}
  \caption{Quantum depletion $\left(\frac{1}{\mathcal{N}}\int G_N(x,x)\,dx\right)$ as a
    function of time (the proportion of non-condensate atoms at
    zero-temperature) simulated with the gapless HFB theory (i.e., using equations(\ref{eq:1dgpe}), (\ref{eq:sigmas}),
    (\ref{eq:ws}), (\ref{eq:factorizeG}), and (\ref{eq:dGdt})) gives a
    depletion proportion that initially scales as $\sim t^2$. }
  \label{fig:non-stationary}
\end{figure}

This formalism now allows an extremely concise representation of the
full dynamical evolution encapsulated in equations~(\ref{eq:eom_GN}) and (\ref{eq:eom_GA});
\begin{equation} i\hbar\pd{G}{t}=\Sigma G - G \Sigma^\dagger
  \label{eq:dGdt}
\end{equation}
where $G$ is defined according to equation~(\ref{eq:factorizeG}). The
consequence of completing the basis by establishing the missing
eigenvector through Gram-Schmidt orthogonalization is now evident. If we
begin with the bare $\Sigma$, as defined in equation~(\ref{eq:simplesigma})
and initialize $G$ to the ground state (meaning $p=0$ and $q=0$) of the corresponding
eigenbasis, then when equation~(\ref{eq:dGdt}) is propagated from this
initial condition, it is evident that the solution is not 
stationary. The number of non-condensate atoms is seen to grow as
$\sim t^2$, as shown in Fig.~\ref{fig:non-stationary}. This implies that we have not in fact determined the
correct ground state. 

This problem arises because, using the language of quantum optics, we are effectively assuming that the
condensate is a coherent field that may act as an infinite classical pump and
can provide a reservoir source for introducing an infinite number of atom-pairs. Furthermore, it does
not cost any energy to introduce a zero-energy quasiparticle within this framework. This is clearly
unphysical for a number of reasons including the fact that, as can be seen in equation~(\ref{eq:eom_phiA}), the
factor of two in front of the interaction between the condensate and
the non-condensate atoms means that it actually costs energy to take
away atoms from the condensate and move them into the non-condensate
fraction, providing the interactions are repulsive (scattering length
positive). There is some literature that suggests simply dropping the
zero-modes entirely to remedy this problem, for example,
Ref.~\cite{FiniteSystems}; however, this generally violates the fundamental 
commutation relations of the bosonic field operator and therefore the
uncertainty principle, so we do not employ that approach here.

We instead employ an alternative solution by including the
second-order terms to generalize the self-energy matrix. This means
that we modify equation~(\ref{eq:sigmas}) to include the effects of the
normal and anomalous densities, and then introduce the renormalization
of the $T$-matrix to give
\begin{eqnarray}
  \Sigma_N &=& -\frac{\hbar^2}{2m}\nabla^2
               - \mu+2V_1[|\phi_a(x)|^2+G_N(x,x)]\,,\nonumber\\
  \Sigma_A &=& V_1[\phi_a(x)^2+G_A(x,x)]\,.
               \label{eq:selfConsistentSigma}
\end{eqnarray}
with both equation~(\ref{eq:simplesigma}) and equation~(\ref{eq:dGdt})
unmodified.  In order to be consistent, however, we must also
generalize the GPE, equation~(\ref{eq:1dgpe}), to
\begin{eqnarray}
  i\hbar\pd{\phi_a(x)}{t}
  &=&-\frac{\hbar^2}{2m}\nabla^2\phi_a(x)
  -\mu\phi_a(x)
  +V_1[|\phi_a(x)|^2
     +2G_N(x,x)]\phi_a(x)\nonumber\\
  &&\quad {}+G_A(x,x)\phi_a^*(x,x) \,.
      \label{eq:gGPE}
\end{eqnarray}
Note that the ground state solution of the GPE is stationary, and thereby determines the value of
the chemical potential that enters the renormalization (see
Section~\ref{sec:renormalization}). Since $G_N(x,x)$ and $G_A(x,x)$
are functionally dependent on the eigenstates themselves, the problem
is nonlinear, and it is necessary to solve the generalized
self-energy, equation~(\ref{eq:selfConsistentSigma}), and the generalized
GPE, equation~(\ref{eq:gGPE}), iteratively until the equations are
self-consistent~\cite{Griffin1996}. We point out that this iterative
process will typically create a small gap in the energy spectrum of
the system around zero energy, and the problem of the unphysical
non-stationary eigensolution that is caused by the zero-energy
subspace is no longer present. The resulting self-consistent solution
is stationary under the evolution given by equation~(\ref{eq:dGdt}) and
provides an accurate ground state initial condition for the subsequent
time-dependent simulations that we present in the rest of the paper.
\begin{figure}
  \centering \includegraphics[width=0.8\textwidth]{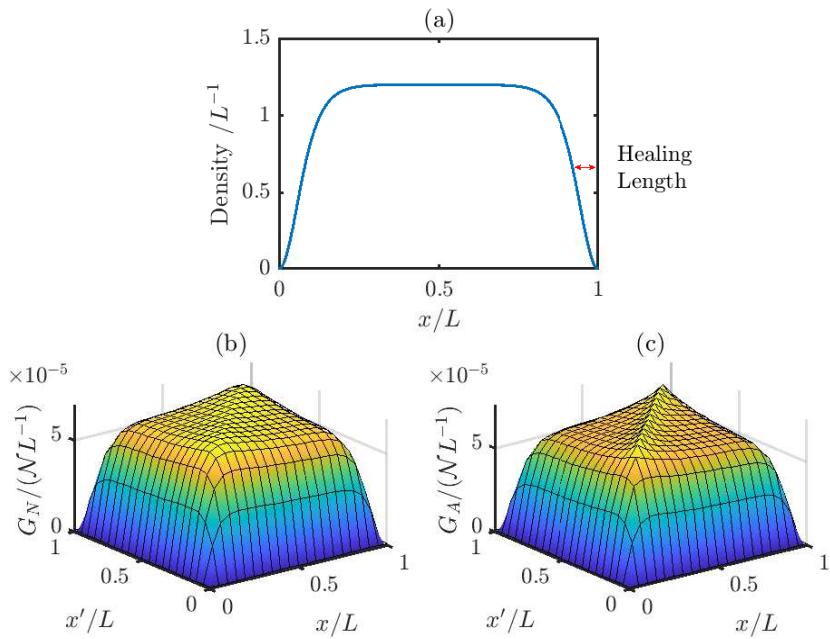}
  \caption{Solutions for a system with total atom number
    $\mathcal{N}=6\times10^{5}$ in a 1D infinite potential well of
    size $L$, with the scattering potential between atoms given by
    $a=10^{-4}~l_\perp^2/L$. (a) Ground
    state condensate density found from the gapped self-consistent
    generalized GPE theory (i.e., replacing equation~(\ref{eq:1dgpe})
    with equation~(\ref{eq:gGPE}) and equation~(\ref{eq:sigmas}) with
    equation~(\ref{eq:selfConsistentSigma})). The length scale over which
    the condensate density falls to zero at the edges of the box is
    known is the healing length. (b) Solution to the normal density
    $G_N(x,x')$ in the ground state as found from the self-consistent
    HFB theory. (c)~Solution to the absolute value of the anomalous
    density $|G_A(x,x')|$ in the ground state as found from the
    self-consistent HFB theory. }
  \label{fig:initialCondition}
\end{figure}

\section{Dynamics of the time-dependent HFB system}
\label{sec:TDHFB}

In the experiment by Clark {\em et al.}~\cite{Firework}, an external sinusoidally oscillating magnetic field is applied, and therefore the scattering length is modulated in the form 
\begin{equation}
   a(t) = a_{\rm dc}+a_{\rm ac}\sin \omega t,
\end{equation}
where $a_{\rm dc}$ is the initial scattering length, and $a_{\rm ac}$
is the amplitude of the oscillating component of the scattering length
at angular frequency $\omega$. 
The dynamics of the system under this modulation is interesting to consider because 
the oscillating external field will inject energy into the
system, and this will result in exciting atoms from the ground state
into higher quasiparticle levels.

We begin our simulations by preparing the system in the
self-consistent ground-state of the HFB theory for a small positive
value of the scattering length using the procedure just described. An
illustration of the resulting condensate, normal and anomalous
densities are shown in figure~\ref{fig:initialCondition}.
After preparing the system in the ground state, we solve
equation~(\ref{eq:simplesigma}) and equation~(\ref{eq:dGdt}) using the
generalized equations~(\ref{eq:selfConsistentSigma}) and (\ref{eq:gGPE})
with a sinusoidal modulation of the scattering potential, i.e.,
\begin{equation}
  V\rightarrow V(t) = V_{\rm dc}+V_{\rm ac}\sin \omega t\,.
\end{equation}

\begin{figure}
\centering
\includegraphics[width=0.85\textwidth]{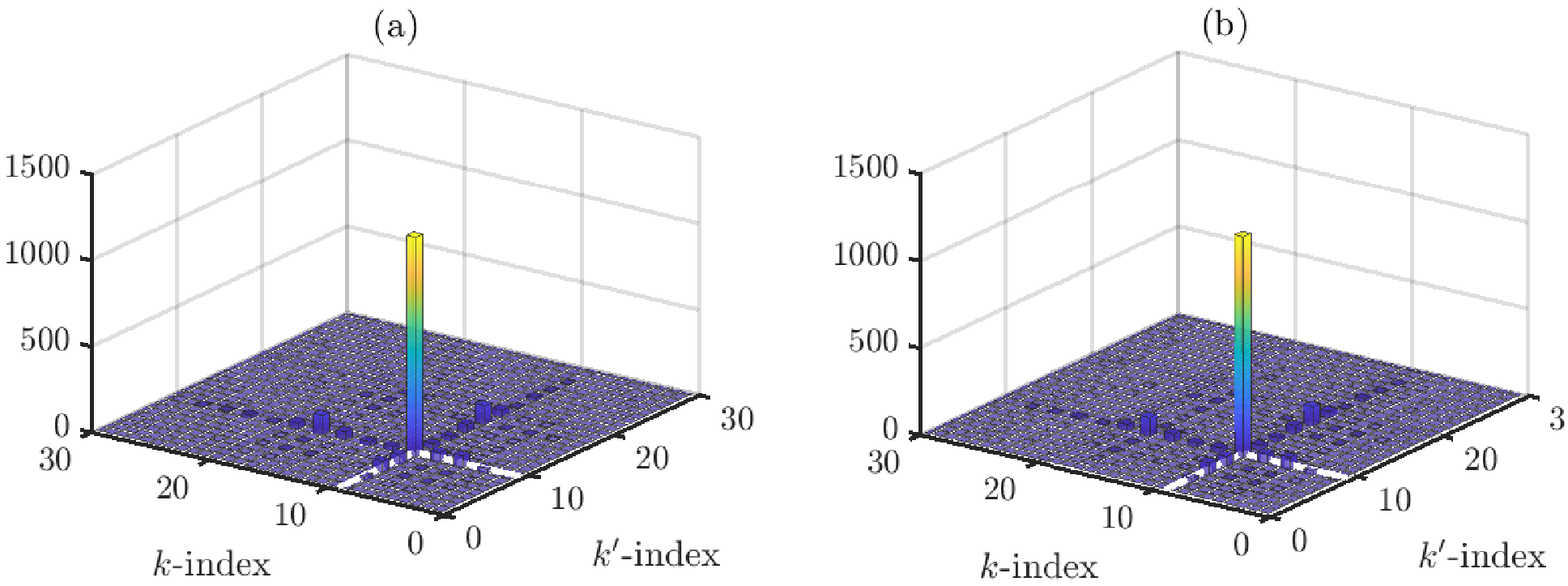}\\
\includegraphics[width=0.4\textwidth]{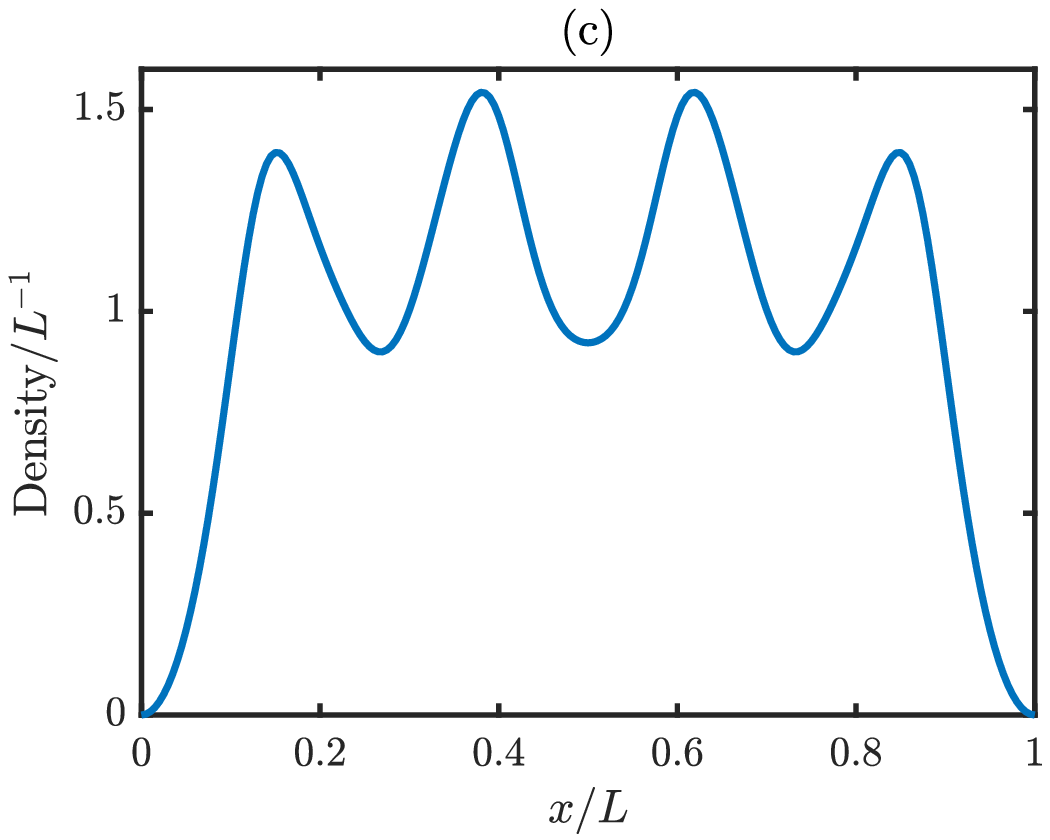}
\caption{Snapshots of (a) absolute value of the matrix elements of the
  upper-left block of the population matrix, $|p|$, and (b) absolute
  value of the matrix elements of the upper-right block of the
  population matrix, $|q|$, at $t=0.1~(mL^2/\hbar)$, starting from the
  initial condition shown in figure~\ref{fig:initialCondition} and
  then continuously driven with amplitude
  $a_\text{ac}=10^{-4}~l_\perp^2/L$ and frequency
  $\omega=1000~(\hbar/mL^2)$. This frequency resonates with the
  quasiparticles with energy $\epsilon_{k^*} = 500~(\hbar^2/mL^2)$,
  corresponding to the resonant wavenumbers shown for reference as
  white lines (at $k$-index $(\pi k)^2\approx 500$). At the resonant
  quasiparticle excitation a clear spike is evident. (c) Density
  profile of the condensate, revealing in general form the spatial
  dependence of the eigenmode function of the resonant quasiparticle
  excitation.}
\label{fig:population}
\end{figure}

In order to interpret our results, we display the occupation
probabilities via the projection of $G(t)$ onto the initial
quasiparticle basis found from the self-consistent HFB Hamiltonian at
time $t=0$. The procedure is as follows. Since the quasiparticle
eigenbasis matrix, $W$, satisfies
$\frac{L}{N}W^\dagger\sigma_zW = \sigma_z$, we may write
\begin{equation}
  W^{-1}=\frac{L}{N}\sigma_zW^\dagger\sigma_z\,.
\end{equation}
Then, according to equation~(\ref{eq:factorizeG}),
\begin{eqnarray}
  \Pi&=&W^{-1}_0G{W^\dagger_0}^{-1}\\ \nonumber
     &=&\frac{L^2}{N^2}\sigma_zW^\dagger_0\sigma_zG\sigma_zW_0\sigma_z
\end{eqnarray}
where $W_0$ is the original self-consistent quasiparticle basis
determined for the initial condition. The resulting population is
shown in figure~\ref{fig:population}.  The height of the peak in the
off-diagonal block (i.e., $q$) is notable since the coherence
saturates the upper bound of the Cauchy-Schwartz inequality,
\begin{equation}
    \left[\delta(x-x')+G_N(x,x)\right]G_N(x',x')\geq |G_A(x,x')|^2,
\end{equation}
which in turn can be interpreted as confirming that the process of
exciting quasiparticles from the condensate is maximally coherent.
The diagonal elements, $p_k$, can be measured by time of flight, since
the quasiparticles transform into regular particles that can be
detected during ballistic expansion. In other words, when the kinetic
energy greatly exceedes the interaction energy, the $k$'s then
effectively label the free momentum, i.e., $k\hbar \pi/L$.
As shown in figure~\ref{fig:population}, when the periodic drive is
turned on continuously for many cycles, essentially only one
quasiparticle mode is resonantly amplified. That is consistent with the narrow spectrum. This physical process can be interpreted as being due, as a consequence of the oscillating drive, to a photon with energy $\hbar\omega$ being absorbed by a pair of atoms, with each of them getting half the energy, $\epsilon_k=\hbar\omega/2$. In addition, the phonon-like collective excitations that correspond to the observed wave-like patterns seen in the condensate density can be interpreted as the
Faraday patterns that typically manifest in different kinds of
parametrically driven fluids~\cite{Faraday}. The pattern resembles the
wavefunction density for the single quasiparticle mode on resonance.
This simulation illustrates that by careful engineering of the drive,
one can potentially prepare a variety of quantum states, selectively
exciting atoms from the condensate field. We now show a few
illustrative examples of interesting cases that employ this technique.

\section{Dynamically Generating Squeezed Quasiparticle States}
\label{sec:squeezing}

A squeezed state refers to a quantum state that has a reduced
uncertainty in one degree of freedom (`squeezed') at the expense of
increased uncertainty in a canonically conjugate
variable~\cite{SpinSqueezing1, SpinSqueezing2, SqueezedLight1,
  SqueezedLight2, AtomInterferometry1, AtomInterferometry2, Shankar_2019}. Such states have been
extensively studied in quantum optics and atomic physics due to their
utility in quantum metrology for producing measurement precision that
exceeds the limits derived from classical states. Here we will show
how to use the resonant quasiparticle excitation in order to generate a
squeezed matter-wave state, anticipating that this could potentially
be applied to quantum matter-wave interferometry.

By driving the system resonantly, we are effectively producing
resonant pairs with well defined energy, and this is reminiscent of
nonlinear optical devices that down-convert pump photons into signal
and idler pairs. Here we will demonstrate that this correspondence is
robust and quantitative by demonstrating how one may calculate the
squeezing parameter associated with the analogous quantity that is regularly computed in the quantum description of light.

In order to do this we assume a weak excitation limit, so that
$G_N(x,x)$ and $G_A(x,x)$ are small compared to $|\phi_a(x)|^2$. Furthermore, we consider the kinetic energy term in the time-dependent GPE to be small, and then we can find a general solution for the condensate that has the form
\begin{equation}
  \phi_a(t) = \phi_0e^{iA\cos(\omega t)}
  = \phi_0\sum_{n=-\infty}^{n=\infty}J_n(A)e^{in\omega t},
\end{equation}
where $A=V_{ac}|\phi_a|^2/\hbar\omega$ and $J_n(\ldots)$ is the Bessel
function of the first kind. We will limit our discussion to the case
of high modulation frequency, in which the photon energy associated
with the drive, $\hbar\omega$, greatly exceeds the mean field shift
associated with the drive amplitude, $V_{ac}|\phi_a|^2$, so that
$A\ll1$. In this case the $n=0$ term completely dominates the series
expansion and we can drop all other terms.

The initial stationary Hamiltonian for the fluctuations can be written as $H_0 = \sum _k\epsilon_k\hat{b}_k^\dagger \hat{b}_k$,  where $\epsilon_k$ is the energy of the $k$-th quasiparticle, and the transformation to a rotating frame involves making the replacement of the quasiparticle operators
\begin{equation}
  \hat{b}_k\rightarrow \hat{b}_ke^{i\epsilon_k t/\hbar}\,.
\end{equation}
The contact interaction term in the Hamiltonian can be derived from the interaction term of equation~(\ref{eq:hamiltonian}),
\begin{equation}
    H_I = \frac{V_{ac}}{2}\sin\omega t\,\int dx\,\hat{\psi}^\dagger(x)\hat{\psi}^\dagger(x)\hat{\psi}(x)\hat{\psi}(x)
\end{equation}
From this point, we keep only the second-order terms in $\delta\hat{\psi}$, because these terms correspond to exponential growth and therefore dominate the solution. In order to simplify the problem further, we assume that the drive frequency $\omega$ corresponds to the
resonance condition $\omega = 2\epsilon_k/\hbar$, and introduce the rotating wave approximation, which
allows us to keep only terms with
$e^{\pm i(\omega - 2\epsilon_k/\hbar)t}$. By representing $\delta\hat{\psi}$ in the quasiparticle basis, we obtain an effective
interaction Hamiltonian
\begin{eqnarray}
    H_I &\approx& \frac{V_{ac}}{2}\sin\omega t\,\int dx\,\left(4|\phi_a|^2\delta\hat{\psi}^\dagger\delta\hat{\psi} + \phi_a^2\delta\hat{\psi}^\dagger\delta\hat{\psi}^\dagger + {\phi_a^*}^2\delta \hat{\psi}\delta\hat{\psi}\right) \nonumber\\
    &\approx&\frac{V_{ac}}{2}\sin\omega t\int dx\bigg[4|\phi_0|^2\nonumber\\
    &&{}\times\sum_{k'}\left(u_{k'}^*\hat{b}_{k'}^\dagger e^{i\epsilon_{k'} t/\hbar}+v_{k'}\hat{b}_{k'}e^{-i\epsilon_{k'}t/\hbar}\right)
    \sum_{k''}\left(u_{k''}\hat{b}_{k''} e^{-i\epsilon_{k''} t/\hbar}+v^*_{k''}\hat{b}_{k''}^\dagger e^{i\epsilon_{k''}t/\hbar}\right)\nonumber\\
    && + \phi_0^2\sum_{k'}\left(u_{k'}^*\hat{b}_{k'}^\dagger e^{i\epsilon_{k'} t/\hbar}+v_{k'}\hat{b}_{k'}e^{-i\epsilon_{k'}t/\hbar}\right)\sum_{k''}\left(u_{k''}^*\hat{b}_{k''}^\dagger e^{i\epsilon_{k''} t/\hbar}+v_{k''}\hat{b}_{k''}e^{-i\epsilon_{k''}t/\hbar}\right) \nonumber\\
    &&+ {\phi_0^*}^2\sum_{k'}\left(u_{k'}\hat{b}_{k'} e^{-i\epsilon_{k'} t/\hbar}+v^*_{k'}\hat{b}_{k'}^\dagger e^{i\epsilon_{k'}t/\hbar}\right)\sum_{k''}\left(u_{k''}\hat{b}_{k''} e^{-i\epsilon_{k''} t/\hbar}+v^*_{k''}\hat{b}_{k''}^\dagger e^{i\epsilon_{k''}t/\hbar}\right)\bigg]\nonumber\\
    &\approx&
    \frac{V_{ac}}{4i}\bigg[\int dx\left(4|\phi_0|^2v_ku_k+\phi_0^2v_k^2+{\phi_0^*}^2u_k^2\right)\hat{b}_k\hat{b}_k \nonumber\\
    &&- \int dx\left(4|\phi_0|^2u_k^*v_k^*+\phi_0^2{u_k^*}^2+{\phi_0^*}^2{v_k^*}^2\right)\hat{b}_k^\dagger \hat{b}_k^\dagger\bigg]
\end{eqnarray}

\begin{figure}
\centering
\includegraphics[width=0.5\textwidth]{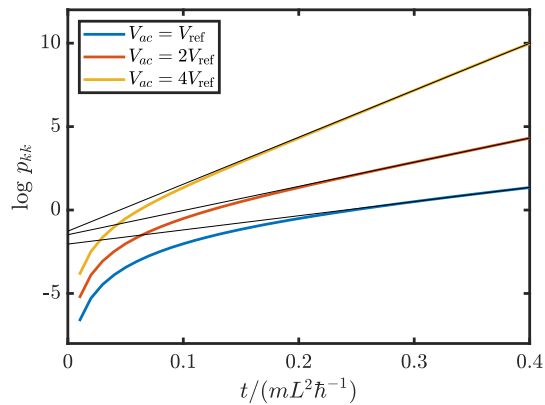}
\caption{Population in
  the resonant quasiparticle mode (labelled $k$) driven with frequency
  $\omega=5500~(\hbar/mL^2)$ as a function of time for different
  modulation amplitudes. The amplitudes are $V_\text{ref}$ (yellow),
  $2V_\text{ref}$ (red), and $4V_\text{ref}$ (blue), where
  $V_\text{ref}=1.25\times10^{-4}~(\hbar^2/mL)$. The slope of $\log~p_{kk}$ at
  large time is equal to twice the squeezing rate, which is
  proportional to the modulation amplitude as shown in
  equation~(\ref{eq:chi}). The squeezing rates calculated from the
  slopes of the curves in the interval $t=[0.3,0.4]~(mL^2/\hbar)$ are
  $4.3,\ 7.2,\ 14.1~(\hbar/mL^2)$, and the squeezing rates
  calculated from equation~(\ref{eq:chi}) are
  $3.6,\ 7.1,\ 14.3~(\hbar/mL^2)$ respectively for amplitudes $V_\text{ref},\ 2V_\text{ref},\ 4V_\text{ref}$.}
  \label{fig:squeezing}
\end{figure}

This corresponds to the interaction Hamiltonian of a parametric
amplifier in nonlinear quantum optics, namely
$H_I = -i\hbar\frac{\chi}{2}(\hat{a}^2 - \hat{a}^{\dagger\,2})$, where $\chi$ represents the second-order nonlinear susceptibility that corresponds to the squeezing rate. We refer to the resulting time-evolved state 
as a squeezed quasiparticle state since the analog is an archetypal
system for creating squeezed states of light. This mapping allows us
to extract the squeezing rate,
 i.e.,
\begin{equation}
  \chi = \frac{V_{ac}}{2\hbar}\int dx\left(4|\phi_a|^2v_ku_k
    +\phi_a^2v_k^2+{\phi_a^*}^2u_k^2\right)\,,
    \label{eq:chi}
\end{equation}
and the squeezing parameter increases with time at this rate, i.e, $\xi=\chi t$.
If we choose the phases of $\phi_a$, $u_k$ and $v_k$ appropriately,
then $\chi$ is real.  As expected from the known optical solutions,
the population in the $k$-th quasiparticle mode grows proportional to
$\sinh^2(\chi t)$. Figure~\ref{fig:squeezing} shows the population as a function of time at different modulation amplitudes. Since $\sinh^2(\chi t)\to e^{2\chi t}/4$ at large $t$, one can extract the squeezing rate from the asymptotic slope of $\log\,p_{kk}$. We confirm that the squeezing rate is proportional to the modulation amplitude, as indicated by equation~(\ref{eq:chi}).

\begin{figure}
\centering
\includegraphics[width=0.5\textwidth]{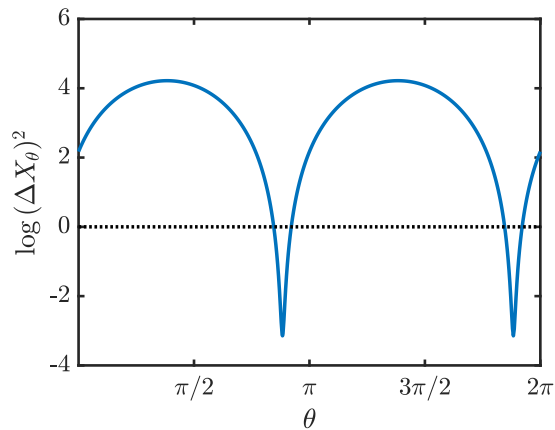}
\caption{The variance of the quadrature as a function of the angle, at $t=0.05~(mL^2/\hbar)$ with modulation amplitude $V_{ac}=2\times10^{-3}~(\hbar^2/mL)$, evaluated using equation~(\ref{eq:variance}). The dotted line is the standard quantum limit, where the variance is equal to~$1$. For a certain range of angles, the variance falls below the standard quantum limit. More specifically, at $\theta=0.88\pi$ the variance has minimum, which means measurements of the quadrature along this direction will have the greatest precision.}
  \label{fig:quadrature}
\end{figure}

Squeezed states are characterized by reduced variance in one quadrature at the expense of increased variance in the other quadrature perpendicular to it. We define the quadrature for the resonant quasiparticles as
\begin{equation}
    X_\theta \equiv \hat{b}_k^\dagger e^{i\theta} + \hat{b}_ke^{-i\theta}\,,
\end{equation}
where $\theta$ is the angle of the orientation of the quadrature.
Then the variance is
\begin{eqnarray}
    \expval{(\Delta X_\theta)^2} &=& \expval{\left(\hat{b}_k^\dagger e^{i\theta} + \hat{b}_ke^{-i\theta}\right)^2}-\expval{\hat{b}_k^\dagger e^{i\theta} + \hat{b}_ke^{-i\theta}}^2 \label{eq:variance}\\ \nonumber 
    &=& 2p_{kk}+1+2Re\{q_{kk}\,e^{-i2\theta}\}\,.
\end{eqnarray}
The variance as a function of $\theta$ is shown in figure~\ref{fig:quadrature}. We see that the variance at certain quadrature phase angles, $\theta$, of states produced by modulation of the scattering potential can fall below the standard quantum limit. The standard quantum limit is the level generated by the uncertainty principle under the assumption that the variance in all angles $\theta$ is uniform.

Although direct measurement of the squeezing may not be as
straightforward to implement as in its optics counterpart, it may be
possible to observe directly the atom coincidence (since the particles
are produced in pairs) on detectors placed in directions corresponding
to opposite momenta, and thereby measure the second-order coherence. The direct analogue of phase sensitive
photodetection (homodyne and heterodyne detection, for example) is
generally more complicated to implement with atoms than light, but in
the next section we propose a possible experiment that could be used
to perform an analogue of such interference measurements on the squeezed
quasiparticle distributions that are generated.

\section{Interferometry with squeezed
  quasiparticles}
\label{sec:interference}

In principle, the diagonal elements of the normal density are the quantities that can be directly
probed with standard atomic density images, for example in dispersive, absorption, or fluorescence imaging techniques. On the other hand,
the off-diagonal elements of the normal density and the anomalous density cannot be directly observed since they are phase
dependent quantities and have complex values that require an
interferometric method to determine the phases. We investigate the phase dependence of the quasiparticle production by analysing two distinct methodologies. One approach is a potential experiment that is capable of performing the phase
measurement through the use of a protocol that is based on the Ramsey
sequence widely used in atomic physics~\cite{Ramsey}. A second alternative approach is closely associated with a recent experiment by Hu {\em
  et al.}  \cite{UnruhRadiation}, who demonstrated that applying a
phase shift to the oscillatory field after driving the system for a
period of time will suppress the non-condensate atom number, and that
a $\pi$ phase shift results in the greatest suppression. 

\begin{figure}
\centering
\includegraphics[width=0.85\textwidth]{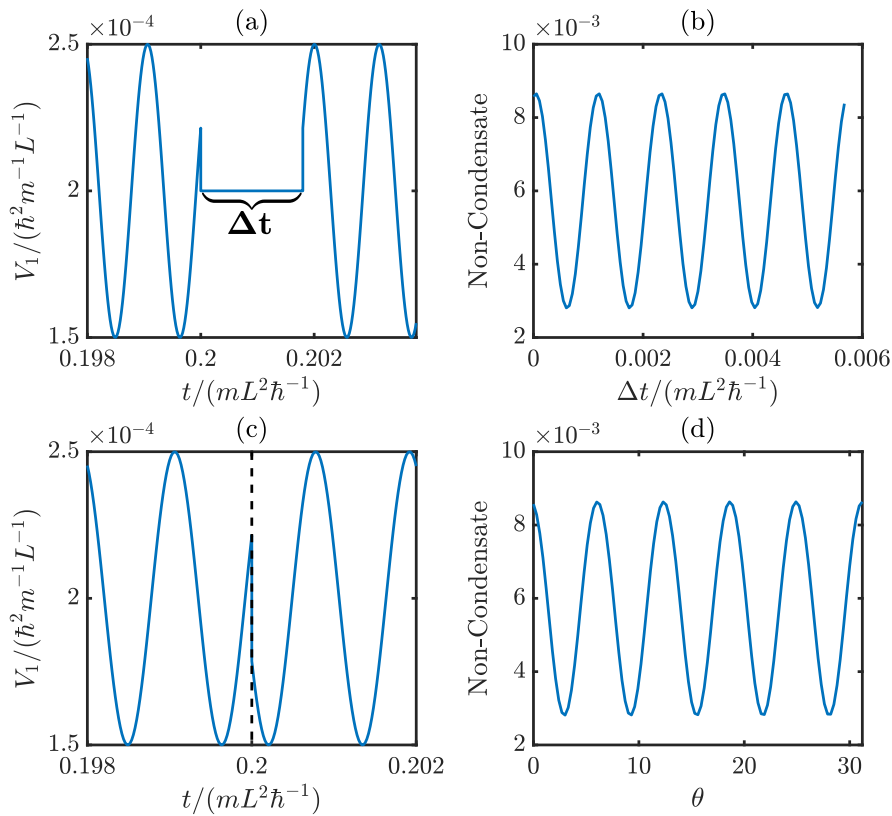}
\caption{The two methods for probing the phase of the quasiparticle
  squeezing.  In both cases, the scattering length at first oscillates
  at frequency $\omega = 5500~(\hbar/mL^2)$ for a period of time $\tau$. At $t=\tau$, in
  (a) the amplitude of the oscillation remains at zero for a time interval
  $\Delta t$, and then the scattering length again oscillates for another
  period of time $\tau$, and in (c) a phase shift $\theta$ is applied to the oscillation
  . Panels (b) and (d) show the resulting
  non-condensate fraction at the final time as a function of $\Delta t$ or phase
  shift $\theta$, respectively, for the two cases.}
\label{fig:Ramsey}
\end{figure}

Our Ramsey protocol is as follows. First, we apply a non-zero $V_\text{ac}$
for a period of time $\tau$ to implement the first oscillatory field
in the Ramsey sequence. We then set $V_\text{ac}$ zero for a brief
waiting period of time $\Delta t$. During this interval the anomalous
density evolves freely at the resonance frequency,
$2\epsilon_k/\hbar$, and because there is no external work done on the
system, the number of non-condensate atoms remains essentially
constant. Next, $V_\text{ac}$ is set to the same nonzero value as
earlier to implement the second oscillatory field, again for the same
period of time $\tau$. This sequence is illustated in
figure~\ref{fig:Ramsey}(a). From our simulation results, shown in
figure~\ref{fig:Ramsey}(b), we observe that the number of
non-condensate atoms oscillates as a function of the free evolution
time, $\Delta t$. This is because a phase difference
$\theta=\omega \Delta t$ accumulates between the anomalous density and
the driving field during the free evolution period. We account for
this behavior by showing that the oscillations observed are a
consequence of the driving field in the second zone either amplifying
or attenuating the anomalous density depending on the accumulated
relative phase.

For comparison, we now examine an abrupt phase change protocol based on the Hu {\em
  et al.}  \cite{UnruhRadiation} experiment. We consider the effect of the phase shift by first modulating the interaction for
a period of time $\tau$, then applying a phase shift $\theta$ to the
oscillating drive, and repeating again the interaction for a period of
time $\tau$, as shown in figure~\ref{fig:Ramsey}(c). The result of the
final non-condensate atom number as a function of the phase shift is
shown in figure~(\ref{fig:Ramsey})(d). It is interesting to compare
this protocol and the resulting fringe pattern to that found from the
first method.  The explanation is that the two methods both operate in a manner that is
analogous to a Mach-Zehnder interferometer, where interference fringes
are seen in the recombination of light propagating along two paths as
the relative accumulated phase is varied. In the first protocol that
we have presented, the phase is accumulated in the anomalous density,
whereas in the second method, a direct phase shift is applied to the
external field. We have observed that both methods result in an
interference pattern with high visibility fringes that allow direct
access for the observer to probe the phase behavior.

The two methods, the complete Ramsey sequence or the abrupt
intermediate phase shift change, can be understood in a similar formalism. Both the phase shift change and the Ramsey
wait-time effectively generate a phase shift in the direction of
squeezing. This manifests as a change in the phase of the squeezing
rate, i.e., $\chi\to \chi e^{i\theta}$, and is associated
with the resonant quasiparticle state evolving under the unitary
operator,
\begin{equation}
  U_\theta(t) = e^{-\frac{\chi}{2}e^{i\theta}(\hat{b}_k^2-\hat{b}_k^{\dagger\,2})t}
\end{equation}
during the subsequent time evolution period. In the Heisenberg picture, the time-evolved
operator $\hat{b}_k$ for the quasiparticle at index $k$ at the end of the
sequence is therefore given by
\begin{eqnarray}
  \hat{b}_k(2\tau) &=& U_\theta^\dagger(\tau)U_0^\dagger(\tau)
                 \hat{b}_k(0)U_0(\tau)U_\theta(\tau)
                 \nonumber\\
  \expval{\hat{b}_k^\dagger(2\tau)\hat{b}_k(2\tau)}
             &=& \cos^2\bigl(\chi\tau\sin\theta\bigr)\sinh^2
                 \bigl(\chi\tau(1+\cos\theta)\bigr)  \nonumber\\
             && \quad {}+\sin^2\bigl(\chi\tau\sin\theta\bigr)
                \cosh^2\bigl(\chi\tau(1+\cos\theta)\bigr).
\end{eqnarray}
Note that at the special point $\theta=\pi$, $U_\pi(t)=U_0(-t)$, and
the second period of modulation simply reverses the effect of the
first period of modulation, so that the final population is
zero. However, we can see that in the numerical simulation, the final
number of non-condensate atoms at a phase shift of $\pi$ is
non-zero. This is because the analytic result is derived using the
rotating-wave approximation, and in the full simulation, the
populations of the off-resonance quasiparticles are not fully reversed
due to the influence of the other terms that were dropped. In this
case, the second period of the modulation may further increase their
populations even at the special point, $\theta=\pi$, leading to the
observed finite non-condensate population. As a consquence, the degree to which the excitations can be fully reversed can be interpreted as a measurement of the fidelity of the protocol for producing quasiparticle squeezing.
fidelity of the preparation of the squeezed quasiparticle state.

\section{Quasi-2D system}
\label{sec:2D}

In order to make a more robust connection with the recent Bose
firework experiment~\cite{Firework}, we would like to generalize the
formalism we have presented from a quasi-1D gas trapped in a box
potential to a quasi-2D gas that is initially trapped by a circular potential with the third out-of-plane
direction frozen. Although this geometry adds new degrees of freedom
to our previous analysis, we may exploit the fact that the circular
system possesses cylindrical symmetry, so that the wavefunction of the
condensate can be solved effectively as a 1D problem in the radial
coordinate. Note that the quasi-2D system differs from the quasi-1D
system in a number of important ways. The momentum correlations will
manifest as angular correlations that may be detected by looking for
atom-atom coincidence on two detectors aligned in opposite
directions. Furthermore, the divergence properties of the
renormalization problem are qualitatively different in two dimensions, as discussed
previously.

We begin by writing the fluctuations in the field operator in the
quasiparticle basis using appropriate indices for two dimensions,
\begin{equation}
  \delta\hat{\psi}(r,\theta) = \sum_{k,l}u_{k,l}(r)e^{il\theta}\hat{b}_{k,l}
  + v_{k,l}^*(r)e^{-il\theta}\hat{b}^\dagger_{k,l}\,,
\end{equation}
where $k$ corresponds to the excitation in the radial coordinate and
$l$ represents the angular momentum quantum number. The angular
momentum will modify the form of the kinetic energy for the 2D
quasiparticles by including a new centrifugal term,
$\hbar^2l^2/2mr^2$, that arises physically from circulation about the
trap center.  Due to cylindrical symmetry, the normal and anomalous
densities should be functions of only three real variables, two radii
and a relative angle, which we denote by $r_1$, $r_2$, and
$\phi\equiv\theta_2-\theta_1$, respectively. For both normal and
anomalous densities, we write the functions in terms of their
expansion in angular momentum,
\begin{equation}
  G_{N,A}(r_1, r_2, \phi) = \sum_lG_{N,A}^{(l)}(r_1, r_2)e^{il\phi}.
\end{equation}    
The time evolution can then be solved by substituting this expansion
into equations~(\ref{eq:eom_phiA})-(\ref{eq:eom_GA}) and using the
appropriate form for the two dimensional kinetic energy.

\begin{figure}
\centering
\includegraphics[width=0.7\textwidth]{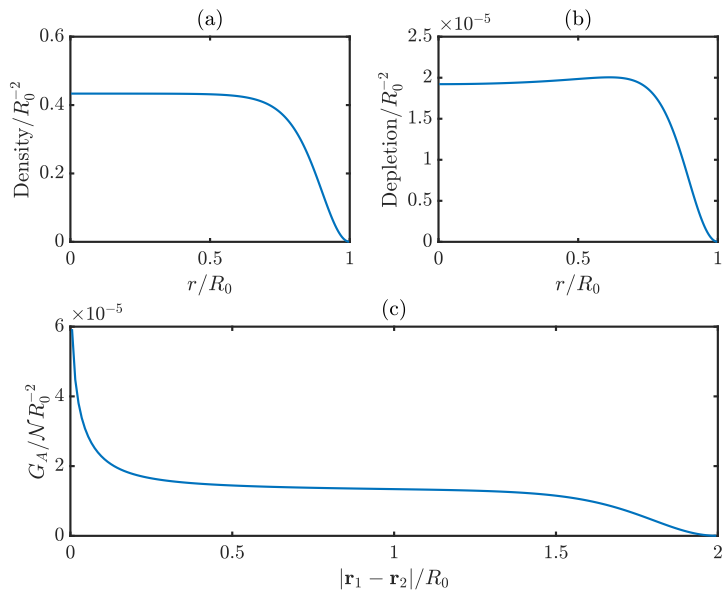}
\caption{Ground state solutions for a system with total atom number
  $\mathcal{N}=6\times10^5$ in an infinite circular box of radius
  $R_0$. The scattering length is set to
  $a=3.99\times10^{-5}~l_\perp$. (a) Condensate density as a function
  of radial position. (b) Quantum depletion density as a function of
  radial position. (c) Anomalous density with the center of mass
  position at the center of the trap 
    i.e., $\vectorsym{r}_1+\vectorsym{r}_2=0$, as a function of the
  relative distance $|\vectorsym{r}_1-\vectorsym{r}_2|$. The
  divergence that scales as the Hankel function of the first kind
  close to the origin is a result of 2D scattering theory.}
\label{fig:2Dground}
\end{figure}

The first case we consider is for the situation in which the circular
trap potential well is infinite and has radius $R_0$,
\begin{equation}
  V_{\rm ext}(r)=\left\{\begin{array}{ll}
                          0&r<R_0\\
                          \infty&\text{otherwise}
                        \end{array}
                      \right.
\end{equation}
We prepare the quantum gas in the ground state with a small repulsive
scattering length~$a$ in order to stabilize the system
mechanically. The repulsive interactions are characterized by the
appropriate 2D $T$-matrix, as discussed in
Section~\ref{sec:renormalization}. Procedurally, we carry out a
similar sequence of steps to those previously discussed for
quasi-1D. First we solve the GPE using imaginary-time propagation, and use that mean-field solution as the first iteration for
the solution of the HFB equations, ignoring the non-condensate terms in the
HFB self-energy. As before, this solution is non-stationary and we
must iterate between the GPE and HFB solutions in order to find a
self-consistent solution whose resulting evolution gives rise to
densities that do not depend on time. The resulting three components,
the condensate, the normal density, and the anomalous density, are
illustrated in figure~\ref{fig:2Dground}.  Note that the anomalous
density diverges in general as the Hankel function of the first kind
as a function of the relative distance
$|\vectorsym{r}_1-\vectorsym{r}_2|$ close to the origin. This is an
analytic result that can be derived by solving the scattering equation, 
equation~(\ref{eq:scattering}), in 2D \cite{Arfken}. This emphasizes an
important point; the anomalous density cannot be accessed directly in
experiment and does not form an observable.

\begin{figure}[]
  \centering \includegraphics[width=0.5\textwidth]{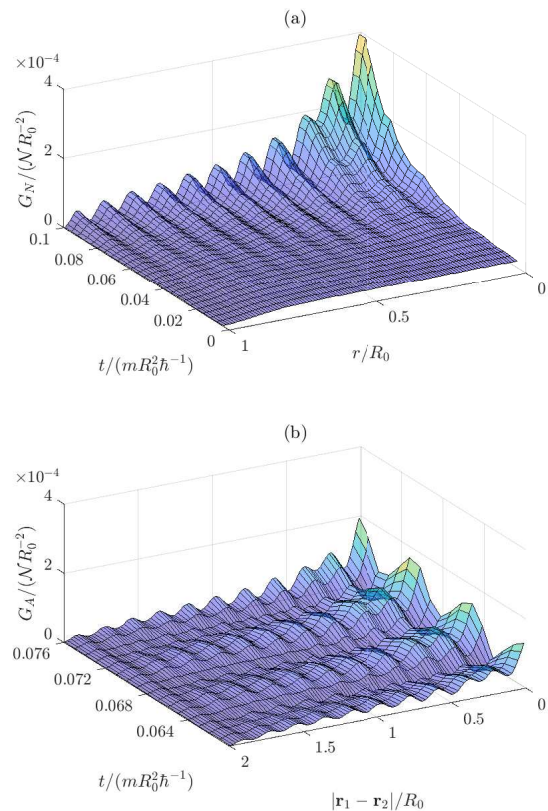}
  \caption{Normal and anomalous densities in a system that is prepared
    in the ground state of the self-consistent HFB solution with a
    positive scattering potential
    $a_\text{dc}=3.99\times10^{-5}~l_\perp$, and then subjected to the
    modulating drive with angular frequency
    $\omega=1200~(\hbar/mR_0^2)$ and constant amplitude
    $a_\text{ac}=3.99\times10^{-5}~l_\perp$. (a) Non-condensate
    density as a function of radial potition
    $r=|\vectorsym{r}_1+\vectorsym{r}_2|/2$ (the origin is on the
    right) and time, i.e.,
    $G_N(r,\, |\vectorsym{r}_1-\vectorsym{r}_2|=0,\,t)$. (b) Magnitude
    of the anomalous density as functions of the relative coordinate
    $|\vectorsym{r_1}-\vectorsym{r_2}|$ (the origin is on the right)
    and time, i.e.,
    $|G_A(r=0,|\vectorsym{r}_1-\vectorsym{r}_2|,t)|$. Only a small
    time interval beginning at $t=0.6~(mR_0^2/\hbar)$ is shown, so
    that the oscillations are resolved.}
    \label{fig:relative}
\end{figure}

Now that we have prepared an accurate initial state, we can then begin
to examine its time evolution when subjected to a drive via a
modulation of the scattering length. Similar to what we saw in
quasi-1D, the modulation leads to excitation of quasiparticles with
energies on resonance with the modulation
frequency. Figure~\ref{fig:relative} shows the normal and anomalous
density as a function of time and relative distance at the center of
the trap. A principal feature of the radial density dependence is the appearance of phonon-like excitations with
well defined wave-number. The non-condensate density increases
monotonically with time, as is consistent with the squeezing picture
discussed earlier. On the other hand the anomalous density oscillates
in time tracking the external field.

\begin{figure}
\centering
\includegraphics[width=1\textwidth]{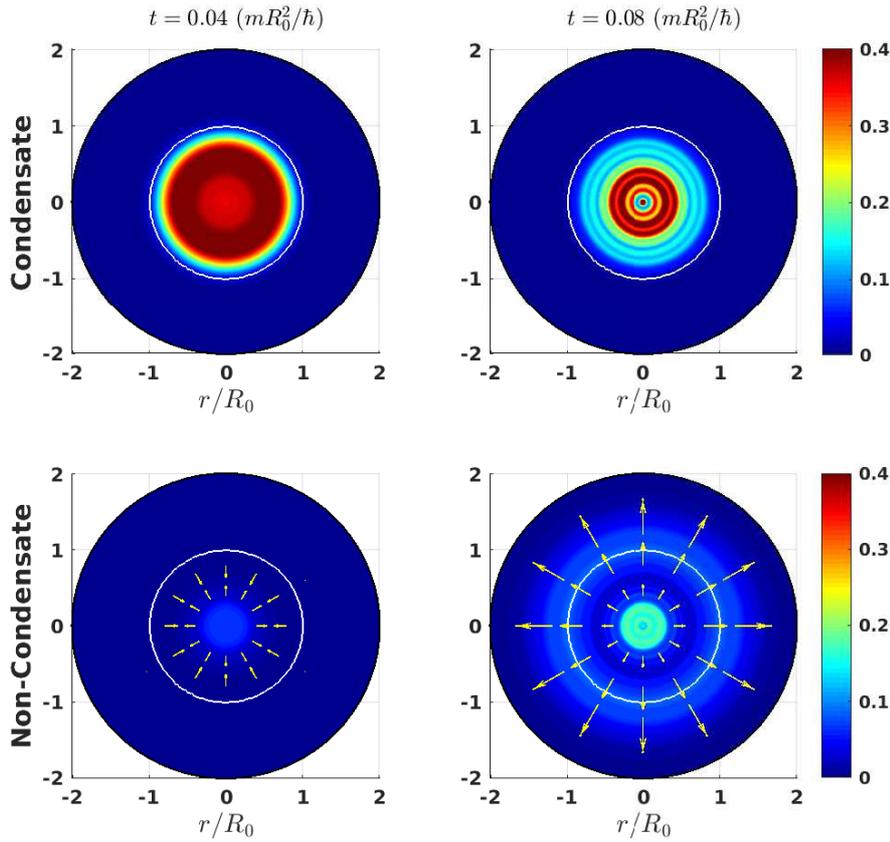}
\caption{Densities of the condensate and the non-condensate at
  $t=0.04$ and $0.08~(mR_0^2/\hbar)$. The dark blue area is the range
  of simulation, and the white line indicates the trapping
  potential. The yellow arrows represent the density currents, with
  their lengths proportional to the magnitude. The system starts with
  the ground state with $a_\text{dc}=1.99\times10^{-5}~l_\perp$ and
  then the scattering length is modulated with amplitude
  $a_\text{ac}=1.99\times10^{-4}~l_\perp$ for all time. The smoothing
  parameter of the finite circular well was set to $\zeta=0.2$. At
  $t=0.04~(mR_0^2/\hbar)$, the atoms in the condensate are pushed
  towards the edge of the trap, but because they are of low energy,
  they cannot escape the trap. The number of high momentum atoms is
  still relatively small, and therefore the density current is also
  small. The density currents at $r=0.4~R_0$ point outwards while
  those at $r=0.8~R_0$ point inwards. Outside the trap, there are few atoms, and the currents are essentially zero. At
  $t=0.08~(mR_0^2/\hbar)$, a large fraction of non-condensate atoms
  with high energy escape the trap, and the density currents near the
  edge of the trap and outside the trap point outwards with large
  magnitudes.
  For clarity, the density currents at $t=0.04~(mR_0^2/\hbar)$ are scaled up 3 times compared to those at $t=0.08~(mR_0^2/\hbar)$. }
\label{fig:densities}
\end{figure}

In order to capture the the dynamics of the high momentum atoms
emitted outwards in the Bose fireworks experiment, we extend our
simulations to a system with a finite trap potential that is higher
than the initial chemical potential but lower than the kinetic energy
of the excited atoms, using a smooth hypertangent functional of form,
\begin{equation}
  V_{\rm ext}(r)=\frac{V_\text{well}}2\left(1+\tanh\bigl(\frac{r-R_0}
    {\zeta R_0}
    \bigr)
  \right)
\end{equation}
where $V_\text{well}$ and $\zeta$ are positive constants. The form of
this external confining potential was chosen to reduce numerical
artifacts. Figure~\ref{fig:densities} shows snapshots of the
condensate and non-condensate densities at time $t=0.04$ and
$0.08~(mR_0^2/\hbar)$ during the drive. At $t=0.04~(mR_0^2/\hbar)$, we
observe that the condensate density is pushed towards the edge of the
trap, and that some non-condensate atoms are generated.  At $t=0.08~(mR_0^2/\hbar)$,
we see phonon-like patterns in the condensate and non-condensate
densities that appear as ripples or waves. We see qualitatively a
residual excited condensate that represents a component that does not
have sufficient energy to overcome the potential barrier, and a
non-condensate density containing much more energetic atoms that is
observed to propagate outwards and leave the finite trap region. These
are due to the energetic quasiparticles created by the drive and form
an experimentally observable quantity in ballistic expansion images.

Figure~(\ref{fig:densities}) also shows the density currents, which
indicate the flow of atoms, including those in the condensate and the
non-condensate. They are computed from
\begin{equation}
    J(r)=-\frac{1}{r}\frac{d}{dt}\int_0^r\rho_\text{tot}(r')\cdot r'\,dr'\,,
\end{equation}
where $\rho_\text{tot}$ is the total density. At
$t=0.04~(mR_0^2/\hbar)$, the number of high momentum atoms is still
relatively small, and therefore the density current is also small. The
density currents at $r=0.4~R_0$ point outwards while those at
$r=0.8~R_0$ point inwards. Outside the trap there are few atoms, and the currents are essentially zero. At
$t=0.08~(mR_0^2/\hbar)$, the density currents near the edge of the
trap and outside the trap point outwards with large magnitudes, which
represents the emission of high momentum atom pairs.

Unlike in the experiment, where jet-like patterns were observed, our
simulation results show isotropic images. This is anticipated since
the functionals we calculate for the condensate, normal, and anomalous
densities, represent probability densities and not individual
realizations (i.e., they are ensemble averages over many experimental realizations.) On the other hand, a given experiment is fundamentally
different in that it represents a single trial that exhibits shot-to-shot
noise associated with the projection that occurs in a single quantum
measurement. In order to model this projection noise it would be
necessary to simulate quantum trajectories~\cite{trajectories}, rather
than solving for the density matrix evolution, and this may be done by
adding white noise to the initial condensate wavefunction~(see for
example Ref.~\cite{DensityWaves}).

\begin{figure}
\centering
\includegraphics[width=0.4\textwidth]{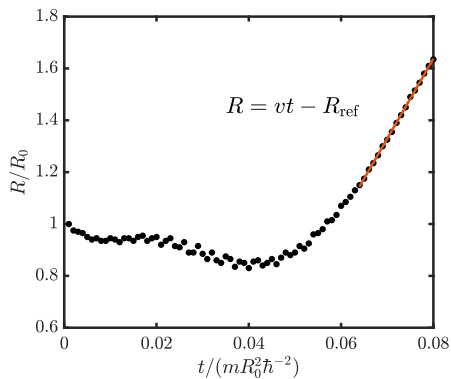}
\caption{The size of the non-condensate as a function of time. The
  size is defined as the radius $R$ that encircles 96.2\% of the
  non-condensate atoms. The system starts with initial chemical potential $\mu=26~(\hbar^2/mR_0^2)$ and is driven
  with modulation frequency $\omega=1000~(\hbar/mR_0^2)$. The speed of the propagation $v$ is given by the slope of the fitted line
  (red), $v=30.8~(\hbar/mR_0)$. This value is 
  approximately equal to $\sqrt{(\hbar \omega-2\mu)/m}$, the speed of a particle whose kinetic energy is half the photon energy minus the chemical potential. We have excluded the data
  points at times when the motion is not ballistic, including at early times where few non-condensate atoms escape the trap
  and the interaction between the condensate and the
  non-condensate is significant.}
  \label{fig:speed}
\end{figure}

In order to demonstrate a quantitative comparison of the energy of the
generated quasiparticles with respect to their ballistic motion, we
present a numerical `time-of-flight' calculation. In this simulation,
we measure the momentum of the atoms in the system by evaluating the
speed at which the gas expands. We define the effective size of the
gas as the radius encircling a large fixed fraction of the
non-condensate (say more than 90\%), so that at $t=0$ we begin with
size $R_0$. Using this metric, from our simulations, we observe that
initially a large amount of non-condensate density is generated close
to the center of the trap and most of the non-condensate atoms have not
left the finite trapping region, so that the size of the gas appears
to be shrinking.  However, later in the evolution and after a
significant fraction of non-condensate atoms escapes the trap, the
expansion of the size of the gas becomes essentially ballistic
(expansion size increasing linearly in time), and the speed of the
expansion is approximately $\sqrt{(\hbar\omega-2\mu)/m}$. 
The reason is as follows. From the HFB equations, we know that
\begin{equation}
    \left(-\frac{\hbar^2}{2m}\nabla^2-\mu+2V|\phi_a|^2\right)u_k +V|\phi_a|^2v_k = \epsilon_ku_k\,.
\end{equation}
Using the approximations $V|\phi_a|^2 \approx \mu$ and $v_k \approx 0$ for large $k$, we may write
\begin{align}
    -\frac{\hbar^2}{2m}\nabla^2u_k \approx \left(\epsilon_k-\mu\right)u_k\,.
\end{align}
Futhermore, since the resonant quasiparticle energy is $\hbar\omega/2$, we find
\begin{align}
    \frac{1}{2}mv^2 \approx \left(\frac{\hbar\omega}{2}-\mu\right)\,.
\end{align}
The speed of the expansion is 
evaluated from the slope of the fitted curve in
figure~\ref{fig:speed}. The fit only includes data points from time
$t=0.065\sim0.08~(mR_0^2/\hbar)$ so that we avoid the initial
transients where interaction energy within the condensate and between
the condensate and noncondensate is significant, and the motion is
not ballistic. The speed we get from the slope agrees well with the analytical result derived above.

\section{\label{sec:conclusion}Conclusion}
In this paper, we have developed a description of a condensate and non-condensate system starting from the many-body field theory Hamiltonian and deriving the evolution equations for the condensate, normal density and anomalous density. 
Since we assumed contact interactions, the contact potential may lead to divergences in the field theory at small and large momenta. We took care of this issue by properly renormalizing the scattering potential.

We solved the quantum fluctuations in the initial stationary state in
1D using the self-consistent HFB theory, which does not involve any free
parameters. 
Then, we simulated the amplification of the quantum
fluctuations with a well-defined energy using the time-dependent
equations. The amplification has aspects similar to the generation of squeezed states of light, and we were able to verify that the variance of the quadrature can fall below the standard quantum limit.
We proposed to observe phase sensitive quantities through two alternate approaches including Ramsey interferometry and discrete phase jumps. We showed how this is able to provide information on the characterization of quasiparticle squeezed states. Finally, we showed simulation results in 2D, and found that the excited
non-condensate atoms eventually leave the trap and propagate outwards at a
well-defined speed, consistent with the experimentally observed 
time-of-flight results. Although we showed simulation results for only quasi-1D and quasi-2D systems, 3D systems would be interesting and can be analyzed systematically using similar approaches.

We have demonstrated a method to generate momentum squeezed states that may be useful for metrology applications. This motivates us to further consider engineering the scattering length as a function of time to generate two-mode squeezed states in quasimomentum that could be injected into matter-wave interferometry. The entanglement properties of such states would be interesting to investigate along with the metrological gain that arises from the quantum advantage.
The importance of pairing in this work also motivates us to consider a similar experiment on fermions, where the interactions could be modulated by variation of the scattering length in the BEC-BCS
crossover regime. The motivation for this is simply that the pairing physics is closely connected with the previously observed fermionic condensation. These considerations will be the subject of future studies.

\section{Acknowledgements}
We thank Athreya Shankar for discussions. This work was supported by NSF PFC Grant No.\ PHY 1734006.
\section*{References}

\providecommand{\newblock}{}

\end{document}